\documentclass[prd,eqsecnum,twocolumn,nofootinbib]{revtex4}
\usepackage{bm}
\usepackage{amssymb}
\usepackage{graphicx}
\usepackage{epstopdf}
\usepackage{mathrsfs}
\usepackage{amsmath}
\usepackage{color}
\numberwithin{equation}{part}

 \usepackage{color}

\begin{document}
\title{
Linearized Stability of Extreme Black Holes}
\author{Lior M.~Burko$^{1}$ and Gaurav Khanna$^2$}
\affiliation{$^1$ School of Science and Technology, Georgia Gwinnett College, Lawrenceville, Georgia 30043 \\
$^2$ Department of Physics, University of Massachusetts Dartmouth, Dartmouth, Massachusetts  02747}
\date{September 28, 2017}
\begin{abstract} 

Extreme black holes have been argued to be unstable, in the sense that under linearized gravitational perturbations of the extreme Kerr spacetime the Weyl scalar $\psi_4$ blows up along their event horizons at very late advanced times. We show numerically, by solving the Teukolsky equation in 2+1D, that all algebraically-independent curvature scalar polynomials approach limits that exist when advanced time along the event horizon approaches infinity. Therefore, the horizons of extreme black holes are stable against linearized gravitational perturbations.
We argue that the divergence of $\psi_4$ is a consequence of the choice of a fixed tetrad, and that in a suitable dynamical tetrad all Weyl scalars, including $\psi_4$, approach their background extreme Kerr values. We make similar conclusions also for the case of scalar field perturbations of extreme Kerr.

\end{abstract}

\maketitle

Black hole (BH) stability has been an important question in the understanding of their physical reality. 
Rigorous analyses have proved linear stability for Schwarzschild BHs for regular initial data  \cite{dafermos-2016}. For the rotating Kerr BH, linear stability has been proved rigorously only for massless scalar field perturbations for the non-extremal case \cite{dafermos-2014}, although mode stability has been demonstrated also for gravitational perturbations \cite{whiting}. 

An interesting class of BHs is that of extreme ones: BHs which have vanishing surface gravity. In classical general relativity
extreme BHs (maximally charged or maximally spinning BHs) behave differently, both physically and mathematically, from non-extremal ones, in a way that draws much attention to them and to nearly-extreme BHs \cite{gralla}. Extreme BHs also play an important role in supersymmetric and string theories, where it is easier to describe them quantum mechanically because of their vanishing surface gravity and consequently vanishing temperature for Hawking radiation \cite{strominger-1996}. 

Recently, it was argued that extreme BHs are unstable: Fields (massless scalar fields or gravitational perturbations) or their transverse derivatives grow unboundedly along their event horizons (EHs). Specifically, Aretakis argued that extreme Reissner-Nordstr\"om BHs are linearly unstable under scalar fields perturbations \cite{aretakis-2010}: Certain transverse derivatives of the time evolution of regular initial data 
grow unboundedly with advanced time. 

Lucietti and Real expanded Aretakis's result also for linearized vacuum gravitational perturbations of extreme Kerr BHs (EK) \cite{lucietti-2012}, and showed that for axisymmetric perturbations  certain second transverse derivatives of the Weyl scalar  $\psi_4$ and certain sixth transverse derivatives of the Weyl scalar  $\psi_0$ blow up in the Hartle-Hawking (HH) tetrad along the EH with advanced time. The HH tetrad is a null tetrad in which the Kinnersley null-tetrad basis vectors ${\bf l},{\bf n}$ are rescaled with the horizon function, so that they are regular on the EH, and specifically, for any finite value of advanced time the Weyl scalars on the EH are finite.   
For non-axisymmetric gravitational perturbations Casals {\it et al} showed that the HH Weyl scalar $\psi_4$ itself blows up along the EH, and that each additional transverse derivative increases the blow-up rate \cite{casals-2016,Sam:PC}, and concluded that spacetime curvature diverged. (Note that it was not claimed in \cite{casals-2016} that curvature scalar invariants blow up. See also \cite{sam-peter}.)
 Lucietti and Real also suggested that when the  full nonlinearity of the Einstein equations is considered, spacetime would evolve such that either a null singularity would evolve instead of an EH, or spacetime would evolve to a non-extreme BH \cite{lucietti-2012}. 

The suggestion that EK are linearly unstable, and that spacetime may evolve a null singularity instead of a regular EH for EK is highly troubling in view of the importance of extreme BHs in both general relativity and string theory. Here, we bring numerical evidence to argue that while the aforementioned results are correct,  the  EK spacetime is nevertheless linearly stable: a scalar polynomial singularity does not evolve. We then argue that in an appropriately chosen tetrad, no instability is seen.



Our approach is to solve the Teukolsky equation \cite{teukolsky-1973} for the Weyl scalars $\psi_4$ and $\psi_0$ in the HH tetrad \cite{teukolsky-press-1974} for EK, using compactified hyperbolical coordinated similar to those used, say, in Ref.~\cite{BKZ}. The major technological innovation we introduced for this study is boundary conditions that allow us to track the evolution of the fields on the EH accurately. 
More specifically, the fields are actually ``evolved'' on the boundary 
(which is the EH in our computational setup) as opposed to computed using 
the boundary conditions in conjunction with data from the ``bulk''. The 
more common approach is to evolve the fields in the bulk, i.e. compute 
the ``right-hand-side'' in the bulk and update the values of the field via 
time-stepping. And then use this evolved data in the bulk along with the 
imposed boundary condition (or a simple extrapolation) to compute the 
fields on the boundary. This approach has the advantage that it is 
computationally cheaper and fairly simple to implement. However, it 
inherently relies on a high degree of smoothness in the solution, thus 
resulting in some inaccuracy in cases wherein a sharp physical feature 
is present at the boundary. Given that that is precisely what is expected 
in this present study, we took the alternate approach of actually evolving 
the fields everywhere, including at the boundary itself. To do this, the right-hand-side is computed at the boundary and the field values are updated at every time-step. Now, computing the right-hand-side involves computing derivatives at the boundary, and that is done using a high-order, one-sided, finite-difference stencil. 
This approach
generated results that were consistent with several of the test cases we
used to validate our computational framework. Detailed results from these 
tests appear later in this paper. 

The numerical scheme we used is presented in detail in Ref.~\cite{ZK-2011} along with 
several stability, convergence and other tests. To summarize the approach: 
(i) The Teukolsky equation, written in hyperboloidal coordinates (based on the 
ingoing Kerr coordinate system) is first cast into a (2+1)D form by separating 
out the axisymmetric $\varphi$ dependence; (ii) The resulting equation is 
rewritten in first-order hyperbolic form; and (iii) A time-explicit, two-step 
Richtmeyer-Lax-Wendroff, second-order finite-difference evolution scheme is 
implemented. We also developed a new fifth-order 
WENO finite-difference scheme \cite{WENO} with third-order Shu-Osher explicit time-stepping \cite{gotlieb-shu-tadmor}. 
This newly implemented method was used to cross check results 
obtained with the second-order code, and to obtain results that were inaccurate with the second-order code.  
The initial data for the evolved fields is specified as a ``truncated'' (in order for it to be compactly supported) Gaussian pulse placed in the strong field, with no support on the horizon. More specifically, in the code's compactified hyperboloidal coordinates ($\rho$, $\tau$) \cite{BKZ}, the Gaussian pulse is centered at $\rho = 5.0M$ and is of width $0.1M$. The truncation window is chosen to be $4.0M$ wide, and centered at the same location as the Gaussian pulse. 

We next find numerically the behavior of the Weyl scalars $\psi_0$ and $\psi_4$ and their $\,\partial_\rho$ gradients  along the EH as functions of advanced time $v$ (``Eddington coordinate"). The gradient $\,\partial_\rho\propto\,\partial_r$, $r$ being the ingoing Kerr radial coordinate. 
We note that $\rho$ is regular on the EH, so that these gradients are effectively gradients with respect to a Kruskal-like coordinate. We further note that as $v\to\infty$ the $\,\partial_\rho$ gradients of $\psi_4,\psi_0$ and also of a scalar field $\phi$ become transverse (i.e., $\,\partial_\rho$ becomes proportional to $\,\partial_u$, $u$ being retarded time), with the relative error at finite late advanced times decaying like $v^{-2}$. For simplicity of discussion, we refer to $\,\partial_\rho$ as a transverse gradient hereafter. 

\begin{figure}
\includegraphics[width=8.5cm]{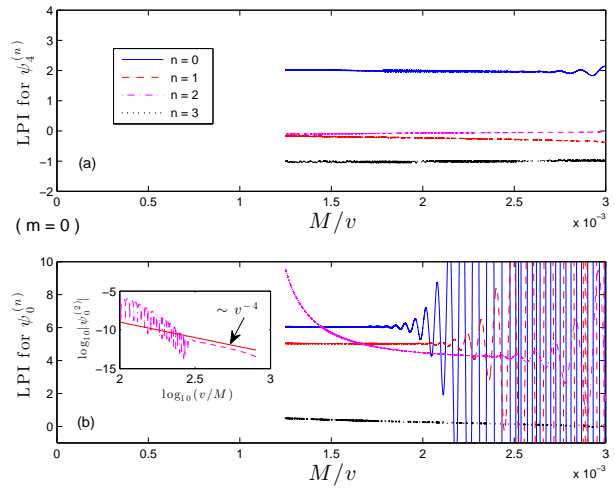}
\caption{The LPIs for the real parts of $\psi_4$ (upper panel - (a)) and $\psi_0$ (lower panel - (b)) and for their first three $\,\partial_\rho$ derivatives along the EH as functions of advanced time, $v$, for axisymmetric ($m=0$) perturbations of EK. The inset in (b) shows $\,\partial^2_\rho \psi_0$ as a function of $v$.  The imaginary parts of the fields behave qualitatively similarly at late times.
Data here and in the figures below are extracted on the surface $\theta=\pi/4$. }
\label{gw_m0_lpi}
\end{figure}

Figure~\ref{gw_m0_lpi} shows the local power indices (LPIs) \cite{burko-ori-97} for the axisymmetric  ($m=0$) case. For the field $\zeta (v)$ we define the LPI $q$ as $q:=-v \zeta_{,v}\zeta^{-1}$. We denote by $\psi_{i}^{(n)}$ the $n^{\rm th}$ transverse derivative of $\psi_i$. 
We find for $m=0$ that for $n=0,1,2,3$ the corresponding $q$ values are $2,0,0,-1$ for $\psi_4$ and $6,5,4,1$ for $\psi_0$. The instability in the field $\psi_4$ is manifest in its third derivative, in accordance with the conclusions of \cite{lucietti-2012}: $q<0$ implies unbounded growth with advanced time along the EH. (We comment that the results here have initial data that are unsupported on the EH. For initial data that are supported on the EH we find results in agreement with \cite{lucietti-2012}.)

\begin{figure}
\includegraphics[width=8.5cm]{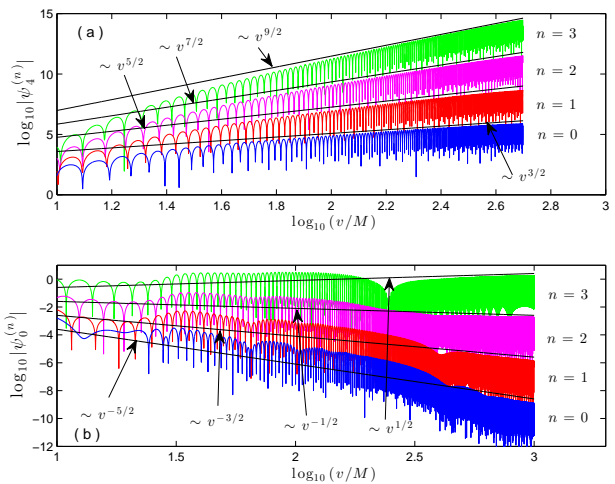}
\caption{The (real parts of the) fields $\psi_4$  (upper panel - (a)) and $\psi_0$ (lower pane; - (b)) and their first three transverse derivatives along the EH as functions of advanced time for non-axisymmetric ($m=2$) perturbations of EK.  In panel (a) we show four reference lines, corresponding to $v^{3/2+n}$, and in panel (b) we show the reference lines for $v^{-5/2+n}$. The imaginary parts behave qualitatively similarly at late times.}
\label{gw_m2}
\end{figure}

In Fig.~\ref{gw_m2} we show the fields $\psi_4$ and $\psi_0$  for the non-axisymmetric case ($m=2$).  Our results for $\psi_4$ are in agreement with the results of \cite{casals-2016} for $n=0,1,2,3$, that is, the late time behavior is found to be $\psi_4^{(n)}(v\gg M)\sim v^{3/2+n}$ and $\psi_0^{(n)}(v\gg M)\sim v^{-5/2+n}$. 

The gravitational case is Ricci flat, and therefore all scalars made with $R$ or $R_{\mu\nu}$ or their derivatives vanish identically. Curvature therefore depends only on the Weyl tensor. A general spacetime in 4D has fourteen algebraically-independent scalars that determine the curvature \cite{petrov}. In vacuum there are only four non-vanishing such scalars, because any curvature invariant can be expressed as a function of a set of the six fundamental (real) invariant eigenvalues of the Weyl tensor. Since the traces of both the Weyl tensor and its dual vanish, there are four independent scalars left \cite{beetle-burko-2002}. 
These scalars may be taken to be the real and imaginary parts of the invariants $I,J$, where $I:={\tilde C}_{\mu\nu\rho\sigma}\,{\tilde C}^{\mu\nu\rho\sigma}$ and 
$J:={\tilde C}_{\mu\nu\rho\sigma}\,{\tilde C}^{\rho\sigma}_{\,\,\alpha\beta}\,{\tilde C}^{\alpha\beta\mu\nu}$, ${\tilde C}_{\mu\nu\rho\sigma}$ being  the self-dual of the Weyl tensor. Our spacetime is even more restricted, because the HH tetrad is a transverse frame ($\psi_1=0=\psi_3$). Since the background is a known EK spacetime, specifically the Weyl scalar $\psi_2$ is known and is constant along the EH, only two algebraically-independent curvature scalars  remain. These scalars can be taken to be the real and imaginary parts of the so-called Beetle-Burko scalar $\xi:=\psi_0\psi_4$ \cite{beetle-burko-2002}. This scalar manifestly contains information only about the radiation field in regions where gravitational radiation is unambiguously defined. 

We show next  that along the EH of EK both the real and the imaginary parts of $\xi$ vanish at late advanced times, so that  $I\to3\psi_2^{\,2}$ and $J\to -\psi_2^{\, 3}$. As $\psi_2$ is that of the background EK, i.e., finite along the EH, both $I$ and $J$ have limits that exist as $v\to\infty$. As $I,J$ exhaust all the algebraically-independent curvature invariants, all scalars made from polynomials in the Weyl tensor have limits that exist as $v\to \infty$. 
We therefore demonstrate that the EK spacetime does not evolve an instability that results in a scalar polynomial singularity. 

\begin{figure}[t]
\includegraphics[width=8.cm]{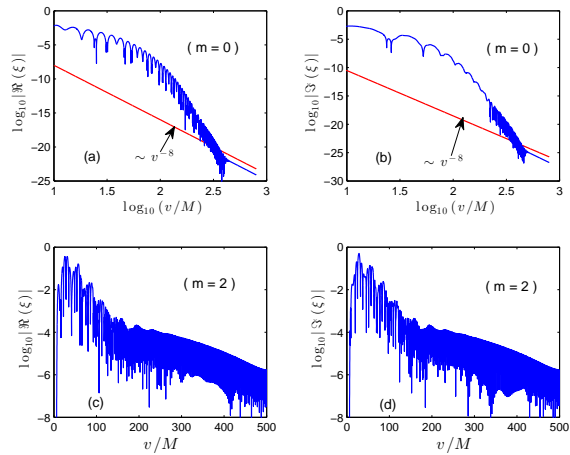}
\caption{The real and imaginary parts of $\xi$ for the axisymmetric ($m=0$) and non-axisymmetric ($m=2$) cases as functions of advanced time along the EH of EK. 
}
\label{curvaturexi}
\end{figure}

Specifically, Fig.~\ref{curvaturexi} shows the real and imaginary parts of $\xi$ for both the axisymmetric and non-axisymmetric cases along the EH of EK as functions of advanced time. We find that in the axisymmetric case  $\Re ( \xi),\Im (\xi )\sim v^{-8}$ for $v\gg M$. In the non-axisymmetric case 
$\Re ( \xi),\Im (\xi )\sim v^{-1}$ for $v\gg M$, so that in either case spacetime curvature along the EH decays to that of the EK background at late advanced times. As argued above, this demonstrates that the EH of the  EK spacetime is indeed stable against linearized gravitational perturbations.

One may ask, how come the blowing up of the Weyl scalar $\psi_4$ does not signify instability, as claimed by \cite{lucietti-2012,casals-2016}. After all, $\psi_4$ is a scalar under coordinate transformations, and therefore all observers would presumably agree on its blowing up. The resolution of this conundrum is that the Weyl scalars are not invariant under transformations of the tetrad vectors. Indeed, under type-III rotations the null tetrad basis vectors ${\bf l }\to A^{-1}{\bf l}$, ${\bf n }\to A^{}{\bf n}$, ${\bf m}\to e^{i\vartheta}{\bf m}$, and $\overline {\bf m}\to e^{i\vartheta}\overline {\bf m}$, where the two real parameters $A,\vartheta$ describe rescaling and rotation, correspondingly, of the tetrad vectors \cite{chandra}. We can choose $\vartheta$ in a way that makes, say, $\Re (\psi_4)=0$, or if we choose another value of $\vartheta$ we can make $\Im (\psi_4)=0$. More importantly, we can choose the rescaling function $A=M/v$, i.e., as our null observer moves along the EH she continuously rescales her tetrad vector ${\bf l}$ linearly in advanced time, and her tetrad vector ${\bf n}$ inversely in advanced time. Correspondingly, $\psi_4\to \psi '_4\sim v^{-2}\psi_4$, and $\psi_0\to \psi'_0\sim v^{2}\psi_0$. Therefore, $\psi '_4\sim v^{-1/2}$ and $\psi '_0\sim v^{-1/2}$ as $v\gg M$. We therefore refer to this dynamical HH tetrad as the symmetric tetrad. We conclude that the blow up of $\psi_4$ in the HH tetrad is a consequence of a problem with the tetrad: If one generalizes the tetrad to a dynamical HH tetrad (``the symmetric tetrad")  in which the basis vectors are continuously rescaled as discussed above, both Weyl scalars $\psi '_4$ and $\psi '_0$ decay to zero. The Beetle-Burko scalar $\xi$ is, however, invariant also under tetrad vector transformations, and therefore is unchanged by this rescaling. Both curvature invariants $I,J$ approach their EK values at late advanced times along the EH. EK are stable because one can find a tetrad in which initially small $\psi '_4, \psi '_0$ remain small along the EH and decay to zero. Notice, that a family of observers, separated by time translations,  who fall into EK and make measurements in the symmetric tetrad are non-parallel-propagated observers. In this family of observers, asymptotically-late daughters see no instability.


Our analysis does not show that any curvature scalar of higher order (i.e., a curvature scalar that includes gradients of the Weyl scalars) blows up along the EH with infinite advanced time. However, we cannot rule out the possibility that curvature scalars of high enough orders do. If that is the case, there would be a non-scalar polynomial singularity (``whimper singularity") evolving, which would be asymptotically delayed \cite{sussman-1988}. Whimper singularities have the feature that in a suitably rotated tetrad the singular behavior disappears. That is, they signify a problem with parallel transport, not a genuine singularity of spacetime, in the sense that there could be (null) observers who are not parallely-propagated, who experience no singular behavior. 



\begin{figure}[t]
\includegraphics[width=8.cm]{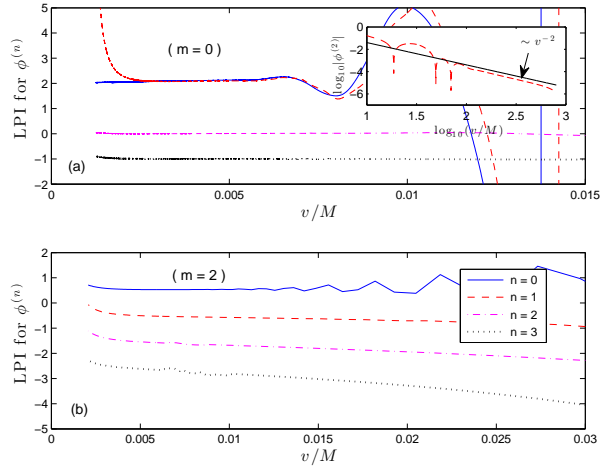}
\caption{The local power indices for a scalar field perturbation for the axisymmetric ($m=0$, upper panel (a)) and non-axisymmetric ($m=2$, lower panel (b)) cases as functions of advanced time along the EH of EK. In the upper panel (a) the inset shows $\,\partial^2_\rho \phi$ as a function of advanced time.
}
\label{scalar}
\end{figure}

Consider next a scalar field. Figure \ref{scalar} shows the LPIs for the axisymemtric ($m=0$) and non-axisymmetric ($m=2$) cases. In both cases we obtain asymptotic LPI values that agree with \cite{casals-2016}. Specifically, in the axisymmetric case we find $q=2,2,0,-1$ and in the non-axisymmetric case $q=1/2,-1/2,-3/2,-5/2$ for $n=0,1,2,3$, respectively. We find that the scalar field itself decays to zero with advanced time, but transverse gradients thereof blow up, consistent with previous results. 

Consider for simplicity an EH null observer on the rotation axis of EK. The gradient of the scalar field, $\,\partial_{\rho}\phi$, blows up for $m=2$ with advanced time. However, observers who use different coordinates disagree on what the gradient is. The only observer-independent way to consider the gradient is to consider a scalar under coordinate transformations. Specifically,  $(\,\nabla_{\alpha}\phi\;\nabla^{\alpha}\phi )^{1/2}\sim (\,\partial_\rho \phi\;\,\partial_v \phi)^{1/2}\sim v^{-1/2}\to 0$ as $v\to\infty$.  Consider next higher-order gradients, say, $\,\nabla_{\alpha_1,\cdots,\alpha_n}\phi$. Also in this case, the scalar $(\,\nabla_{\alpha_1,\cdots,\alpha_n}\phi\;\nabla^{\alpha_1,\cdots,\alpha_n}\phi)^{1/2}\sim v^{-1/2}$ vanishes at infinite advanced time. 

We cannot calculate the perturbations of the Riemann tensor in the scalar field case, as we have a fixed Kerr background. However, we can use the (linearized) Einstein equations to find the Ricci tensor: We write the Einstein equations as $R_{\mu\nu}=8\pi\, (T_{\mu\nu}-Tg_{\mu\nu}/2)$.  We can then calculate the scalar field energy-momentum tensor from the scalar field perturbation $\phi$, 
$T^{\mu\nu}[\phi ]=(g^{\mu\alpha}g^{\nu\beta}+g^{\mu\beta}g^{\nu\alpha}-g^{\mu\nu}g^{\alpha\beta})\,\partial_{\alpha}\phi\,\partial_{\beta}\phi$.  
The Ricci scalar $R\sim (\,\partial_\rho \phi \; \partial_v \phi)\sim v^{-1}\to 0$ as $v\to\infty$ for $m=2$. The curvature scalar $R_{\mu\nu}R^{\mu\nu}\sim (\,\partial_\rho \phi\;\,\partial_v \phi)^2\sim v^{-2}$ in the non-axisymmetric case. We conjecture that all other curvature scalar polynomials made with $R$ and $R_{\mu\nu}$ are also well behaved as $v\to\infty$ along the EH. We cannot, however, find a compete set of algebraically-independent scalar polynomials as we did in the gravitational case. 

We next examine scalars constructed from gradients of $R$ and $R_{\mu\nu}$.  Consider $\nabla^{\mu}R\,\nabla_{\mu}R$. Comparing with $R^2$, we now introduce one additional $\partial_\rho$ and one additional $\partial_v$ derivatives. The effect of both tends to cancel each other, and this scalar behaves like $v^{-2}$ in the non-axisymmetric case. Also scalars such as $\,\nabla^{\sigma}R^{\mu\nu}\,\nabla_{\sigma}R_{\mu\nu}\sim v^{-2}$ and $R^{\mu \nu}\,\nabla_{\mu}R\,\nabla_{\nu}R\sim v^{-3}$. We did not find a scalar made with derivatives of the curvature that does not decay to 0 as $v\to\infty$. We propose that in the case of a scalar field, neither a scalar polynomial singularity nor a non-scalar polynomial one evolves. 


The authors thank Stefanos Aretakis, Sam Gralla, Amos Ori, and Andy Strominger (and his research group) for discussions. 
GK acknowledges research support from the National Science Foundation (award no. PHY-1701284) and Air Force Research Laboratory (agreement no. 10-RI-CRADA-09). Many of the computations for this work were performed on systems provided by Microway Inc. and Nvidia.

\end{document}